# Selectively controlled ferromagnets by electric fields in van der Waals ferromagnetic heterojunctions

Zi-Ao Wang[1,2,*], Weishan Xue[3,*], Faguang Yan[1], Wenkai Zhu[1], Yi Liu[3], Xinhui Zhang[1,2], Zhongming Wei[1,2,4], Kai Chang[1,2,4], Zhe Yuan[3,†], Kaiyou Wang[1,2,4,‡]

[1] *State Key Laboratory of Superlattices and Microstructures, Institute of Semiconductors, Chinese Academy of Sciences, Beijing 100083, China*

[2] *Center of Materials Science and Optoelectronics Engineering, University of Chinese Academy of Sciences, Beijing 100049, China*

[3] *Center for Advanced Quantum Studies and Department of Physics, Beijing Normal University, Beijing 100875, China*

[4] *Beijing Academy of Quantum Information Sciences, Beijing 100193, China*

**Charge transfer plays a key role at the interfaces of heterostructures, which can affect electronic structures and ultimately the physical properties of the materials. However, charge transfer is difficult to manipulate externally once the interface formed. Here, we report electrically tunable charge transfer in $Fe_3GeTe_2/Cr_2Ge_2Te_6/Fe_3GeTe_2$ all-magnetic van der Waals heterostructures, which can be exploited to selectively modify the magnetic properties of the top or bottom $Fe_3GeTe_2$ electrodes. The directional charge transfer from metallic $Fe_3GeTe_2$ to semiconducting $Cr_2Ge_2Te_6$ remarkably modifies magnetic anisotropy energy of $Fe_3GeTe_2$, leading to the dramatically suppressed coercivity. The electrically selective control of ferromagnets demonstrated in this study could stimulate the development of spintronic devices based on van der Waals magnets.**

Interfacial charge transfer can trigger chemical reactions in organic materials consisting of functional groups [1-5] and modulate the physical properties of the constituent materials in heterostructures [6-16], owing to their distinctive electronic structures. It has been widely observed in many heterostructures, such as oxide heterostructures [9,10], topological insulator heterostructures [11], metal-ferromagnet heterostructures [12,13] superconductor-ferromagnet heterostructures [14,15] and ferromagnet heterostructures [16], and the resulting proximity effects can strongly modify the physical properties of the heterostructures. For instance, the Curie temperature of a ferromagnetic semiconductor can be elevated by attaching it to a ferromagnetic metal [16]. It was recently found that the magnitude of charge transfer in heterostructures could be modified via the insertion of dielectric spacers or changing the component ratio [17,18]. However, charge transfer and its effects are difficult to alter once heterostructures form. In particular, electrically tunable charge transfer in heterostructures has not been achieved, which could open new avenues for designing novel functional devices.

Electronic structure modification by electron or hole transfer is usually short-ranged owing to the finite extension of electronic wave functions across an interface. Therefore, atomically sharp interfaces of two-dimensional van der Waals (vdW) magnetic materials are suitable for exploring the electrical manipulation of magnetic properties via charge transfer in heterostructures [19-21]. A vast range of vdW magnetic materials have been discovered to date, including ferromagnetic metals [22], ferromagnetic semiconductors [23,24], antiferromagnetic semiconductors [25] and antiferromagnetic topological insulators [26]. Among these materials, Fe$_3$GeTe$_2$ (FGT) is a typical vdW ferromagnetic metal with strong perpendicular magnetic anisotropy and a high Curie temperature $T_C$ of ~ 220 K [27]. The vdW ferromagnetic semiconductor Cr$_2$Ge$_2$Te$_6$ (CGT) exhibits perpendicular magnetic anisotropy with a small coercivity of ~ 3.4 mT below 65 K (the $T_C$ of bulk CGT) [28]. We investigate the charge transfer effect in vdW ferromagnet heterojunctions of CGT (which has a very low conductivity) and metallic FGT.

The CGT/FGT bilayers are fabricated using mechanical exfoliation of CGT and FGT crystals (HQ graphene, [29]) and dry transfer technique (details in Supplementary S0). The magnetic properties of a typical CGT/FGT bilayer (~ 5 nm/17 nm) are probed by measuring the anomalous Hall effect [Fig. 1(a)]. To determine the effect of the presence of CGT, an FGT Hall device of the same thickness is studied as a reference [Fig. 1(b)]. The linearity of current-voltage (*I-V*) characteristic curves for both devices [Fig. S1(a)] confirms the ohmic contact between FGT and Pt electrodes. The near-insulating behavior of CGT (Fig. S3) results in current mainly flowing through FGT layer. Metallic conduction of both devices is evidenced by the decreasing longitudinal resistances ($R_{xx}$) as temperature decreases [Figs. S1(b) and S1(c)]. The two devices exhibit very different anomalous Hall resistance hysteresis loops at 10 K under out-of-plane magnetic fields ($H_\perp$). Figure 1(c) shows a remarkable suppression of the coercivity ($H_C$) of the CGT/FGT bilayer compared with that of FGT device. To further verify this, we fabricated three other devices with different thicknesses of FGT (~ 10, 15 and 20 nm), where the FGT layers were half covered with similar CGT (5 ~ 6 nm).

The well suppressed $H_C$ of the bilayer part is observed compared with the part only having FGT in all these devices (Fig. S5).

Further analysis of the hysteresis loops at different temperatures ($T$) for both devices (Fig. S2) yields the temperature-dependent coercivity summarized in Fig. 1(d). At low temperatures (stage I: $T < 50$ K), the coercivity of individual FGT of 205 mT is suppressed to 75 mT for the CGT/FGT bilayer, and the coercivities of both devices decrease with increasing temperature, consistent with previously reported results for FGT [30]. At stage II (50 K - 90 K), the $H_C$ of CGT/FGT bilayer rapidly increases with the temperature. Above 90 K (stage III: 90 K - 230 K), the CGT/FGT and FGT curves almost coincide and decrease with increasing temperature. This reduction of the $H_C$ gradually disappears at stage II, near the $T_C$ of CGT ~ 65 K, suggesting the FGT coercivity is related to the ferromagnetic (FM) phase of neighboring CGT. In addition, the turning point of $H_C$ from stage I to stage II coincides with the minimum longitudinal resistance [Figs. S1(b) and S5], suggesting that the suppression of $H_C$ may originate from changes in electronic structure.

To elucidate the mechanism for coercivity reduction in CGT/FGT bilayers, we perform first-principles calculations for a bilayer consisting of three-layers CGT and two-layers FGT [CGT(3)/FGT(2)], as shown in Fig. 2(a), where the magnetizations of CGT and FGT are considered to be parallel and antiparallel to each other. The calculated energy difference between the parallel and antiparallel configurations is below 1 meV per atom (Fig. S7), which is smaller than the thermal fluctuation even at low temperatures ($3k_BT/2 = 1.3$ meV at $T = 10$ K). To verify this, we measured the magnetic hysteresis loops by perpendicular magneto-optical Kerr effect (PMOKE) for CGT, FGT and CGT/FGT bilayer, respectively, see details in Supplementary S5. The bilayer exhibits two separate hysteresis loops for FGT and CGT. Both the calculation and PMOKE measurements suggest negligible magnetic coupling between CGT and FGT.

Since the electronic structures of semiconducting CGT and metallic FGT are significantly different, contact between CGT and FGT results in the redistribution of electrons near the interface and charge transfer across the interface. Figure 2(b) shows the calculated density of states (DOS) of individual three-layer CGT and two-layer FGT. Using the vacuum energy as a common reference, FGT has a higher Fermi energy than the conduction band bottom of ferromagnetic CGT, such that the transfer of electrons from FGT to CGT upon contact creates an interfacial dipole, as confirmed by the directly calculated DOS of the CGT(3)/FGT(2) bilayer. The layer-resolved DOS is shown in Fig. 2(c), and the first CGT atomic layer at the interface becomes conductive with partially occupied conduction bands. The calculated spontaneous charge transfer only occurs at the interface layers, where the FGT layer loses 0.05 electrons per unit cell to the neighboring CGT layer. The occupation of any other layer only changes less than 0.01 electrons confirming that the CGT(3)/FGT(2) bilayer is representative for our study without loss of generality. This metallized CGT atomic layer allows conduction electrons transport across the FGT/CGT interface and hence causes more scattering, which results in a larger resistance of the CGT/FGT bilayer below 50 K, as shown in Figs. S1(b) and S5.

The loss of electrons changes the magnetic properties of FGT. Figure 2(d) shows the calculated magnetic anisotropy energy (MAE) for FGT with different thicknesses as a function of the change in valence electron density. The MAE is defined as the total energy difference between the magnetization within and perpendicular to the two-dimensional plane. For any thickness we calculated, the MAE decreases with increasing density of lost electrons but tends to saturate as the number of electrons gained increases. Therefore, charge transfer from FGT to CGT in the bilayer effectively reduces the electron density in FGT, which in turn reduces the MAE. The reduced MAE results in coercivity suppression for FGT in contact with CGT at low temperatures, as shown in Fig. 1(d). Our calculation shows that the magnetic moments of FGT (Fig. S10) are nearly unaffected (less than 11%) by the reduced density of valence electrons in agreement with literature [22]. The negligible coupling between CGT and FGT suggests that the suppressed coercivity of FGT in CGT/FGT bilayer shown in Figs. 1(c) and 1(d) indeed reflects the large modulation in the MAE of FGT. Therefore, the magnetism of FGT can be manipulated significantly by either raising [22] or decreasing the electron density.

Above $T_C$ (~ 65 K), CGT loses long-range magnetic order, and the magnetic moments of the Cr atoms become randomly oriented. We use a disordered localized moment model to simulate this paramagnetic (PM) phase of CGT, and the calculated DOS is shown by the orange line in Fig. 2(b). The energy of the conduction band bottom is higher for the PM CGT than the FM phase. Charge transfer from the neighboring FGT to the PM CGT can still occur, but the number of transferred electrons decreases. Consequently, the decrease in the MAE of FGT due to electron loss becomes weaker when the attached CGT transitions to the PM phase above $T_C$, which is in agreement with the experimentally observed gradual recovery of the FGT coercivity at stage II (50 K - 90 K).

Charge transfer occurs spontaneously upon placing FGT on top of CGT. The resulting interface dipole pointing from FGT to CGT essentially lowers the energy barrier to confine the valence electrons in FGT (Fig. S11). If a bias voltage is applied to the FGT/CGT/FGT all-magnetic tunnel junction, the effective electrical field can enhance or reduce the interface dipole depending on the field direction. To simulate this scenario, we use a physically transparent jellium slab to model a metallic FGT layer and artificially introduce an interface dipole between the metal and tunnel barrier. Figure 2(e) shows the calculated change in the electron density of the right metallic layer as a function of the applied voltage, where the left electrode is always grounded. Under a positive bias, the enhanced dipole at the right interface significantly increases the loss of electrons (Fig. S11). The magnitude of the electron transfer depends nonlinearly on the external voltage, corresponding to a varying capacitance, which is the signature of a mesoscopically confined electronic system [31]. Note that the interface dipole due to the spontaneous charge transfer at the CGT/FGT interface plays an important role in this process, without which electron transfer would be considerably reduced. Figures 2(d) and 2(e) show that the coercivity of FGT can be effectively tuned by external fields.

To investigate the tunability of charge transfer by electrical fields, we fabricated a magnetic tunneling junction (MTJ) with CGT (~ 6 nm) sandwiched by two FGT electrodes of different thicknesses (9.7 nm and 14.5 nm) to guarantee different $H_C$ (FGT$_{top}$ was thicker than FGT$_{bottom}$), as shown in Fig. 3(a). The magneto-transport properties of the junction were measured by the two-terminal method (details in Supplementary S0). Figure 3(b) shows the measured nonlinear *I-V* curve, which confirms the tunneling characteristic. We further measured the resistance of the device with scanning the external perpendicular magnetic field under different fixed DC voltages at a low temperature (10 K, unless specified otherwise). The measured hysteresis loops of the resistance (*R*) plotted in Fig. 3(c) exhibit typical spin-valve behavior with two resistance states under negative voltages. The high- and low-resistance states represent the resistance at parallel (↑↑ or ↓↓) and antiparallel magnetizations of two FGT electrodes (↑↓ or ↓↑), respectively, see details in Supplementary S8. Sharp jumps of resistances occur at the switching fields corresponding to the FGT electrodes coercivities. The negative magnetoresistance is originated from the opposite spin polarization of the two FGT electrodes [32] (see Supplementary S9).

Figure 3(d) shows the calculated differential resistances (d$R$/d$H_\perp$) under different bias voltages ($V_{bias}$) form the measured magnetoresistance curves (details in Supplementary S10), and the positions of the minimum and maximum d$R$/d$H_\perp$, indicated by the grey dotted lines, are used to quantitatively determine the switching fields. At $V_{bias}$ = -475 mV, the switching fields are 90 mT and 170 mT. The switching fields are 40 mT/170 mT at $V_{bias}$ = -500 mV and 0 mT/170 mT at $V_{bias}$ = -600 mV. The applied negative voltage corresponds to the electrical field pointing toward the bottom

FGT electrode, which enhances the interface dipole at the bottom CGT/FGT interface but suppresses the one at the top FGT/CGT interface, as indicated by the large and small arrows in Fig. 3(a), respectively. The lowered electron density in the bottom FGT electrode significantly reduces the anisotropy field, leading to a dramatically reduced $H_C \sim 0$ mT at $V_{bias}$ = -600 mV. The unchanged switching magnetic field of 170 mT for the top FGT electrode is attributed to the saturation of MAE with increasing electron density. This experimental observation is consistent with the theoretically predicted tunability of the FGT MAE under the application of an external voltage.

The physical picture presented above is confirmed by measuring the magneto-transport of the MTJ under opposite electric fields. Additional measurements show that the application of positive electric fields reduces the coercivity of the top FGT electrode without affecting the bottom FGT electrode (Fig. S14). For comparison, Fig. 4(a) shows the measured spin-valve curves for voltages of the same magnitude but opposite signs obtained by sweeping the magnetic fields, where two-resistance states are observed in both cases. The switching fields can be well distinguished in the corresponding $dR/dH_\perp$ curves shown in Fig. 4(b). At $V_{bias}$ = -800 mV, the switching fields for the top and bottom FGT electrodes of 170 mT and 0 mT, respectively, are exactly the same as those at $V_{bias}$ = -600 mV. However, at $V_{bias}$ = +800 mV, the switching field becomes 0 mT for the top FGT electrode and 120 mT for the bottom FGT electrode.

The measured $H_C$ for the top and bottom FGT electrodes under positive and negative bias voltages is summarized in Fig. 4(c). Under a negative $V_{bias}$, electron transfer at the top FGT/CGT interface is suppressed, and the $H_C$ of top FGT electrode is nearly constant, corresponding to its initial value. The bottom FGT electrode loses electrons due to enhanced electron transfer at the bottom CGT/FGT interface, and therefore, the reduced MAE results in a decrease of $H_C$, which becomes zero at $V_{bias}$ = -600 mV indicating that the magnetic anisotropy of bottom FGT electrode switches from perpendicular to in-plane, corresponding to the negative MAE in Fig. 2(d). This behavior is reversed under a positive $V_{bias}$: the $H_C$ of top FGT electrode decreases to zero at $V_{bias}$ = +800 mV, and the $H_C$ of bottom FGT electrode tends to saturate to its initial value. The nonlinear change in measured $H_C$ is in agreement with that predicted from the MAE calculations and charge transfer analysis [Figs. 2(d) and 2(e)].

To summarize, we demonstrate that the magnetic properties of the top or bottom ferromagnetic electrodes in vdW ferromagnetic heterojunctions can be selectively controlled using voltage-manipulated directional charge transfer. Our work not only provides a new possibility to manipulate the magnetic properties, but also points to the promising potential for utilizing vdW magnets in novel electrically controllable spintronic devices.

This work was supported by the National Key R&D Program of China (Grant No. 2017YFA0303400), the Beijing Natural Science Foundation Key Program (Grant No. Z190007), the National Natural Science Foundation of China (Grant No. 61774144, 62005265, 11734004 and 12174028), and the Chinese Academy of Science (Grant No. QYZDY-SSW-JSC020, XDB44000000 and XDB28000000).

[†] Corresponding author.

zyuan@bnu.edu.cn

‡ Corresponding author.

kywang@semi.ac.cn

\* These authors contributed equally to this work.

[1]  H. Alves, A. S. Molinari, H. Xie, and A. F. Morpurgo, Nat. Mater. **7**, 574 (2008).

[2]  T.-C. Tseng, C. Urban, Y. Wang, R. Otero, S. L. Tait, M. Alcamí, D. Écija, M. Trelka, J.M. Gallego, N. Lin *et al*., Nat. Chem. **2**, 374 (2010).

[3]  K. Vandewal, S. Albrecht, E. T. Hoke, K. R. Graham, J. Widmer, J. D. Douglas, M. Schubert, W. R. Mateker, J. T. Bloking, G. F. Burkhard *et al*., Nat. Mater. **13**, 63 (2014).

[4]  P. B. Deotare, W. Chang, E. Hontz, D. N. Congreve, L. Shi, P. D. Reusswig, B. Modtland, M. E. Bahlke, C. K. Lee, A. P. Willard *et al*., Nat. Mater. **14**, 1130 (2015).

[5]  S. Rafiq, B. Fu, B. Kudisch, and G. D. Scholes, Nat. Chem. **13**, 70 (2021).

[6]  M. N. Grisolia, J. Varignon, G. Sanchez-Santolino, A. Arora, S. Valencia, M. Varela, R. Abrudan, E. Weschke, E. Schierle, J. E. Rault, J.-P. Rueff, A. Barthélémy, J. Santamaria, and M. Bibes, Nat. Phys. **12**, 484 (2016).

[7]  T. P. Lyons, D. Gillard, A. Molina-Sánchez, A. Misra, F. Withers, P. S. Keatley, A. Kozikov, T. Taniguchi, K. Watanabe, K. S. Novoselov, J. Fernández-Rossier, and A. I. Tartakovskii, Nat. Commun. **11**, 6021 (2020).

[8]  D. Zhong, K. L. Seyler, X. Linpeng, N. P. Wilson, T. Taniguchi, K. Watanabe, M. A. McGuire, K.-M. C. Fu, D. Xiao, W. Yao, and X. Xu, Nat. Nanotechnol. **15**, 187 (2020).

[9]  J. A. Bert, B. Kalisky, C. Bell, M. Kim, Y. Hikita, H. Y. Hwang, and K. A. Moler, Nat. Phys. **7**, 767 (2011).

[10] X. R. Wang, C. J. Li, W. M. Lü, T. R. Paudel, D. P. Leusink, M. Hoek, N. Poccia, A. Vailionis, T. Venkatesan, J. M. D. Coey *et al*., Science **349**, 716 (2015).

[11] F. Wang, X. Wang, Y.-F. Zhao, D. Xiao, L.-J. Zhou, W. Liu, Z. Zhang, W. Zhao, M. H. W. Chan, N. Samarth *et al*., Nat. Commun. **12**, 79 (2021).

[12] Y. Tserkovnyak, A. Brataas, and G. E. W. Bauer, Phys. Rev. B **66**, 224403 (2002).

[13] H. K. Gweon, H.-J. Park, K.-W. Kim, K.-J. Lee, and S. H. Lim, NPG Asia Mater. **12**, 23 (2020).

[14] J. Chakhalian, J. W. Freeland, G. Srajer, J. Strempfer, G. Khaliullin, J. C. Cezar, T. Charlton, R. Dalgliesh, C. Bernhard, G. Cristiani, H.-U. Habermeier, and B. Keimer, Nat. Phys. **2**, 244 (2006).

[15] T. Y. Chien, L. F. Kourkoutis, J. Chakhalian, B. Gray, M. Kareev, N. P. Guisinger, D. A. Muller, and J. W. Freeland, Nat. Commun. **4**, 2336 (2013).

[16] C. Song, M. Sperl, M. Utz, M. Ciorga, G. Woltersdorf, D. Schuh, D. Bougeard, C. H. Back, and D. Weiss, Phys. Rev. Lett. **107**, 056601 (2011).

[17] J. Nichols, X. Gao, S. Lee, T. L. Meyer, J. W. Freeland, V. Lauter, D. Yi, J. Liu, D. Haskel, J. R. Petrie *et al*., Nat. Commun. **7**, 12721 (2016).

[18] Y. Zhang, K. Shinokita, K. Watanabe, T. Taniguchi, M. Goto, D. Kan, Y. Shimakawa, Y. Moritomo, T. Nishihara, Y. Miyauchi, and K. Matsuda, Adv. Mater. **32**, 2003501 (2020).

[19]   K. S. Burch, D. Mandrus, and J.-G. Park, Nature (London) **563**, 47 (2018).

[20]   C. Gong and X. Zhang, Science **363**, eaav4450 (2019).

[21]   M. Gibertini, M. Koperski, A. F. Morpurgo, and K. S. Novoselov, Nat. Nanotechnol. **14**, 408 (2019).

[22]   Y. Deng, Y. Yu, Y. Song, J. Zhang, N. Z. Wang, Z. Sun, Y. Yi, Y. Z. Wu, S. Wu, J. Zhu, J. Wang, X. H. Chen, and Y. Zhang, Nature (London) **563**, 94 (2018).

[23]   C. Gong, L. Li, Z. Li, H. Ji, A. Stern, Y. Xia, T. Cao, W. Bao, C. Wang, Y. Wang *et al.*, Nature (London) **546**, 265 (2017).

[24]   B. Huang, G. Clark, E. Navarro-Moratalla, D. R. Klein, R. Cheng, K. L. Seyler, D. Zhong, E. Schmidgall, M. A. McGuire, D. H. Cobden *et al.*, Nature (London) **546**, 270 (2017).

[25]   Z. Wang, M. Gibertini, D. Dumcenco, T. Taniguchi, K. Watanabe, E. Giannini, and A. F. Morpurgo, Nat. Nanotechnol. **14**, 1116 (2019).

[26]   M. M. Otrokov, I. I. Klimovskikh, H. Bentmann, D. Estyunin, A. Zeugner, Z. S. Aliev, S. Gaß, A. U. B. Wolter, A. V. Koroleva, A. M. Shikin *et al.*, Nature (London) **576**, 416 (2019).

[27]   A. F. May, S. Calder, C. Cantoni, H. Cao, and M. A. McGuire, Phys. Rev. B **93**, 014411 (2016).

[28]   V. Carteaux, D. Brunet, G. Ouvrard, and G. Andre, J. Phys. Condens. Matter **7**, 69 (1995).

[29]   See information for HQ graphene at http://www.hqgraphene.com.

[30]   C. Hu, D. Zhang, F. Yan, Y. Li, Q. Lv, W. Zhu, Z. Wei, K. Chang, and K. Wang,

Sci. Bull. **65**, 1072 (2020).

[31] S. E. Nigg and M. Büttiker, Phys. Rev. Lett. **102**, 236801 (2009).

[32] M. Sharma, S. X. Wang, and J. H. Nickel, Phys. Rev. Lett. **82**, 616 (1999).

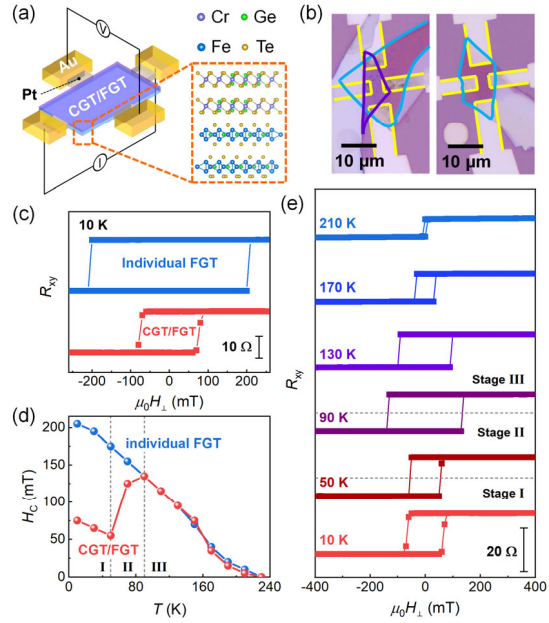

**FIG. 1.** Suppression of $H_C$ in CGT/FGT bilayer. (a) Schematic of CGT (top)/FGT (bottom) bilayer and Hall measurement apparatus. (b) Optical images of the CGT/FGT bilayer (left) and individual FGT device (right). The thin parts of CGT and FGT flakes are outlined in purple and blue, respectively; and the edges of Pt electrodes are outlined in yellow. (c) Out-of-plane magnetic-field-dependent Hall resistances ($R_{xy}$) of the individual FGT device (top panel) and CGT/FGT bilayer (bottom panel) at 10 K. (d) Extracted coercivities of the CGT/FGT bilayer and individual FGT device at different temperatures. (e) Temperature-dependent anomalous Hall resistance hysteresis loops of the CGT/FGT bilayer.

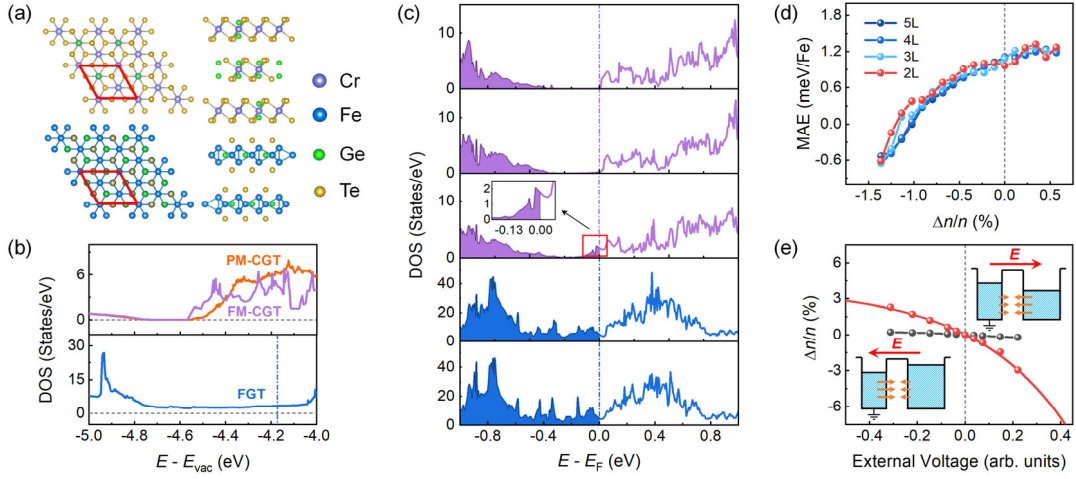

**FIG. 2** Mechanism for charge transfer in CGT/FGT heterostructures. (a) Schematic of the lattice structure of a CGT(3)/FGT(2) bilayer consisting of three-layers CGT and two-layers FGT. In the two-dimensional plane, a $1 \times 1$ CGT unit cell matches a $\sqrt{3} \times \sqrt{3}$ FGT cell by better than 1.3%, as shown by red rhombic frames. (b) Calculated DOS of individual FGT (2) and CGT (3) in FM (purple line) and PM (orange line) phases. (c) Layer-resolved DOS of the CGT(3)/FGT(2) bilayer. The dash-dotted line denotes the Fermi energy. The inset shows the partially occupied conduction bands in the interfacial CGT layer. (d) Calculated MAE of FGT with different thicknesses as a function of the change in valence electron density. A positive MAE indicates perpendicular magnetic anisotropy. (e) Calculated electron density with (red symbols) and without (grey symbols) interface dipoles for a jellium thin film as a function of the applied external voltage. The left electrode is always kept grounded. The insets illustrate the interface dipoles (orange arrows) variation depending on the applied electric field.

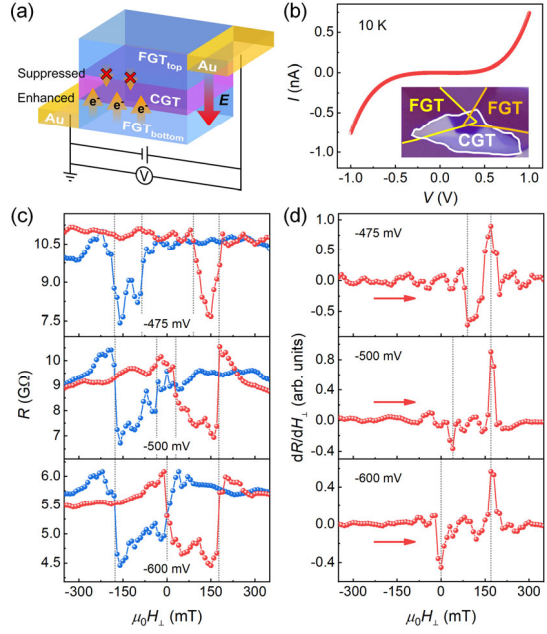

**FIG. 3** Manipulation of charge transfer by electric fields. (a) Schematic of an FGT/CGT/FGT MTJ and the effect of electric fields on charge transfer. (b) $I$-$V$ curve for the FGT/CGT/FGT MTJ at 10 K. The inset is the optical image of this FGT/CGT/FGT MTJ. (c) Out-of-plane magnetic-field-dependent resistance of the FGT/CGT/FGT MTJ under various $V_{bias}$ ranging from -475 mV to -600 mV at 10 K, after subtracting the noise background. The grey lines indicate the positions of switching fields. (d) Calculated $dR/dH_\perp$ curves under different negative $V_{bias}$. The grey lines indicate the positions of the minimum and maximum.

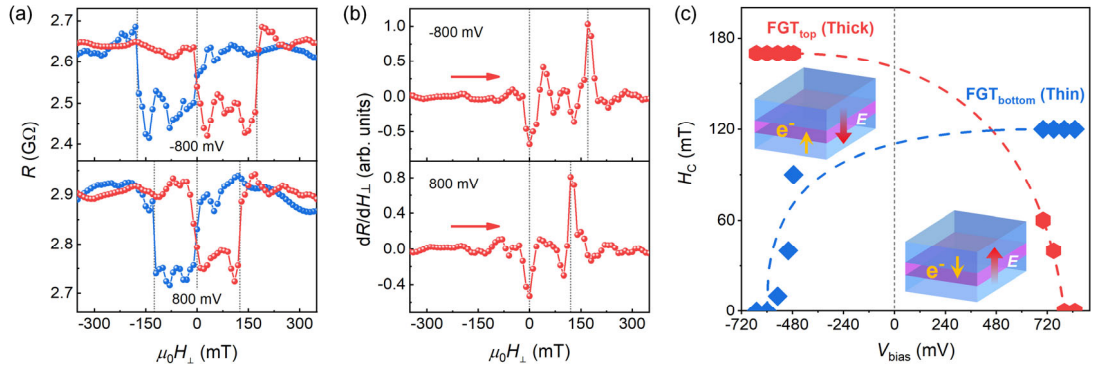

**FIG. 4** Selective control of $H_C$ by applying electric fields. (a) Magnetoresistance curves obtained under opposite bias voltages at 10 K, after subtracting the noise background. The grey lines indicate the positions of switching fields. (b) Calculated $dR/dH_\perp$ curves under opposite $V_{bias}$. The grey lines indicate the positions of the minimum and maximum. (c) Extracted bias-voltage-dependent $H_C$ of the top and bottom FGT electrodes. The dashed lines are to guide the eye. The insets illustrate the electron transfer under the electric fields in opposite directions.